\documentclass[11pt]{article}
\usepackage{coling2020}
\usepackage{times}
\usepackage{url}
\usepackage{latexsym}

\usepackage{booktabs}       
\usepackage{amsfonts}       
\usepackage{nicefrac}       

\usepackage{amsmath, amssymb, amsthm}
\usepackage{multirow}
\usepackage{algorithmic}
\usepackage{algorithm}
\usepackage{graphicx}

\usepackage{enumitem}

\usepackage{microtype}

\colingfinalcopy 

\title{Neural Fair Collaborative Filtering}
\author{ Rashidul Islam,$^1$ Kamrun Naher Keya,$^1$ Ziqian Zeng,$^2$ Shimei Pan,$^1$ {and}  James Foulds$^1$\\ 
$^1$University of Maryland, Baltimore County, USA\\
$^2$Hong Kong University of Science and Technology\\
\small\{kkeya1, islam.rashidul, shimei, jfoulds\}@umbc.edu, zzengae@cse.ust.hk 
}

\begin{document}
\maketitle
\begin{abstract}
A growing proportion of human interactions are digitized on social media platforms and subjected to algorithmic decision-making, and it has become increasingly important to ensure fair treatment from these algorithms. In this work, we investigate gender bias in collaborative-filtering recommender systems trained on social media data. We develop neural fair collaborative filtering (NFCF), a practical framework for mitigating gender bias in recommending sensitive items (e.g. jobs, academic concentrations, or courses of study) using a pre-training and fine-tuning approach to neural collaborative filtering, augmented with bias correction techniques. We show the utility of our methods for gender de-biased career and college major recommendations on the MovieLens dataset and a Facebook dataset, respectively, and achieve better performance and fairer behavior than several state-of-the-art models.
\end{abstract}

\section{Introduction}
Machine learning (ML) systems are increasingly used to automate decisions with direct impact on our daily lives such as 
credit scoring, loan assessments, crime prediction, hiring, and college admissions. There is increasing awareness that ML algorithms can affect people in unfair ways with legal or ethical consequences when used to automate decisions~\cite{barocas2016big,angwin2016machine}, for example, exhibiting discrimination 
towards certain demographic groups. 
As social media platforms are a major contributor to the number of automated data-driven decisions that we as individuals are subjected to, it is clear that such ML fairness issues in social media can potentially cause substantial societal harm. Recommender systems are the primary method for a variety of ML tasks for social media data, e.g. suggesting targets of advertisements, products (e.g. movies, music etc.), friends, web pages, and potentially consequential suggestions such as romantic partners or even career paths. Despite the practical challenges due to labor market dynamics~\cite{kenthapadi2017personalized}, professional networking site-based job recommendation approaches~\cite{bastian2014linkedin,frid2019find,gutierrez2019explaining} are helpful for job seekers and employers. However, biases inherent in social media data can lead recommender systems to produce unfair suggestions. For example, XING, a job platform similar to LinkedIn, was found to rank less qualified male candidates higher than more qualified female candidates~\cite{lahoti2019ifair}. 

Recommendations in educational and career choices are another important motivating application for fair recommender systems.    
Students' academic choices can have significant impacts on their future careers and lives. 
In 2010, women accounted for only 18\% of the bachelor’s degrees awarded in computer science~\cite{broad2014recruiting}, and interventions to help bridge this gap are crucial ~\cite{beede2011women}.  Recommender systems can reinforce this disparity, or ---potentially--- help to mitigate it. 

In this paper, we investigate gender bias in recommender systems trained on social media data for suggesting sensitive items (e.g. academic concentrations or career paths).  For social media data in particular, we typically have abundant implicit feedback for user preferences of various \emph{non-sensitive} items in which gender disparities are acceptable, or even desirable (e.g. ``liked'' Facebook pages, movies, or music), but limited data on the \emph{sensitive} items (e.g., users typically have only one or two college concentrations or occupations).  User embeddings learned from the non-sensitive data can help predict the sparse sensitive items, but 
may encode harmful stereotypes, as has been observed for word embeddings~\cite{bolukbasi2016man}.  Furthermore, the distribution of sensitive items typically introduces further unwanted bias due to societal disparities in academic concentrations and career paths, e.g. from the ``leaky pipeline'' in STEM education \cite{beede2011women}. 

We propose a practical technique to mitigate gender bias in sensitive item recommendations while resolving the above challenges. 
Our approach, which we call \emph{neural fair collaborative filtering (NFCF)}, achieves accurate predictions while addressing \emph{\textbf{sensitive data sparsity}} by pre-training a deep neural network on big implicit feedback data for non-sensitive items, and then fine-tuning the neural network for sensitive item recommendations.  We perform two bias corrections, to address  \emph{(1) bias in the \textbf{input embeddings} due to the non-sensitive items}, and \emph{(2) bias in the \textbf{prediction outputs} due to the sensitive items}.  An ablation study shows that \emph{\textbf{both} interventions are important for fairness}.   
We demonstrate the utility of our method on two datasets: \emph{MovieLens} (non-sensitive \emph{movie ratings} and sensitive \emph{occupations}), and a  \emph{Facebook} dataset (non-sensitive \emph{Facebook page ``likes''} and sensitive \emph{college majors}). 
%
Our main  contributions include:
\begin{itemize}
    \item 
    We propose a pre-training + fine-tuning neural network method for fair recommendations on social media data.
    \item We 
    propose two de-biasing methods: 1) de-biasing latent embeddings, and 2) learning with a fairness penalty.
    \item 
    We perform extensive experiments showing both fairness and accuracy benefits over baselines on two datasets.
\end{itemize}

\section{Background and Related Work}
In this section we formalize the problem, and discuss collaborative filtering with implicit data, and fairness metrics.

\subsection{Problem Formulation}
Let $M$ and $N$ denote the number of users and items, respectively. Suppose we are given a user-item interaction matrix $\textbf{Y}\in \mathbb{R}^{M\times N}$ of \emph{implicit feedback} from users, defined as
\begin{equation}
y_{ui} = 
    \begin{cases}
    1, & \text{if $u$ interacts with $i$}\\
    0, & \text{otherwise.}
    \end{cases}
\end{equation}
Here, $y_{ui}=1$ when there is an interaction between user $u$ and item $i$, e.g. when $u$ ``likes'' Facebook page $i$. In this setting, a value of $0$ does not necessarily mean $u$ is not interested in $i$, as it can be that the user is not yet aware of it, or has not yet interacted with it. 
While interacted entries reflects users' interest in items, the unobserved entries may just be missing data. Therefore, there is a natural scarcity of strong negative feedback.

The collaborative filtering (CF) problem with implicit feedback is formulated as the problem of predicting scores of unobserved entries, which can be used for ranking the items. 
The CF model outputs $\hat{y}_{ui}=f(u,i|\Theta)$, where $\hat{y}_{ui}$ denotes the estimated score of interaction $y_{ui}$, $\Theta$ denotes model parameters, and $f$ denotes the function that maps model parameters to the estimated score. If we constrain $\hat{y}_{ui}$ in the range of [0,1] and interpret it as the probability of an interaction, we can learn $\Theta$ by minimizing the following negative log-likelihood objective function:
\begin{equation}\label{eq:loss}
    L = - \sum_{(u,i)\in \chi \cup \chi^{-}} y_{ui}\log\hat{y}_{ui} + (1-y_{ui})\log(1-\hat{y}_{ui})\mbox{ ,} 
\end{equation}
where $\chi$ represents the set of interacted user-item pairs, and $\chi^{-}$ represents the set of negative instances, which can be all (or a sample of) unobserved interactions.

Learning with implicit feedback becomes more challenging when there is not enough observed interaction data per user. In our setting, we further suppose that items $i$ are divided into non-sensitive items ($i_n$) and sensitive items ($i_s$). For example, the $i_n$'s can be \emph{Facebook pages} where user preferences may reasonably be influenced by the protected attribute such as gender, and the user's ``likes'' of the pages are the implicit feedback. Since each user $u$ can (and often does) ``like'' many pages, $u$'s observed non-sensitive data ($u$-$i_n$) is typically large. On the other hand, $i_s$ may be the users' \emph{occupation} or \emph{academic concentration} provided in their social media profiles.  We desire that the recommendations of $i_s$ to new users should be unrelated to the users' gender (or other protected attribute). Since each user $u$ may typically be associated with only a single occupation (or other sensitive personal data rarely disclosed), the data sparsity in the observed sensitive item interactions ($u$-$i_s$) is a major challenge.   
As a result, it is difficult to directly predict $u$-$i_s$ interactions based on other $u$-$i_s$ interactions. Typical collaborative filtering methods can suffer from overfitting in this scenario, and overfitting often amplifies unfairness or bias in the data such as harmful stereotypes \cite{zhao2017men,foulds2018intersectional}. Alternatively, the non-sensitive interactions $u$-$i_n$ can be leveraged, but these will by definition encode biases that are unwanted for predicting the sensitive items. 
For example, liking the \emph{Barbie doll} Facebook page may be correlated with being female and negatively correlated with \emph{computer science}, thus implicitly encoding societal bias in the career recommendations.

\subsection{Neural Collaborative Filtering}
Traditional matrix factorization (MF) models~\cite{koren2009matrix} map both users and items to a joint latent factor space of dimensionality $v$ such that user-item interactions are modeled as inner products in that space.  Each item $i$ and user $u$ are associated with a vector $q_i\in R^{v}$ and $p_u\in R^{v}$, with
\begin{equation}\label{eq:MF}
    \hat{y}_{ui} = q_{i}^{T}p_{u}+\mu + b_i + b_u\mbox{ ,}
\end{equation}
where $\mu$ is the overall average rating, and $b_u$ and $b_i$ indicate the deviations of user $u$ and item $i$ from $\mu$, respectively. 

Neural collaborative filtering (NCF)~\cite{he2017neural} replaces the inner products in matrix factorization with a deep neural network (DNN) 
that learns the user-item interactions.  
In the input layer, the users and items are 
typically one-hot encoded, then 
mapped into the latent space with an embedding layer.  
NCF combines the latent features of users ${p}_u$ and items ${q}_i$ by concatenating them.  
Complex non-linear interactions are modeled by stacking hidden layers on the concatenated vector, e.g. using a standard multi-layer perceptron (MLP).  
A commonly used architecture is a tower pattern, where the bottom layer is the widest and each successive layer has a smaller number of neurons~\cite{he2016deep}.
\subsection{Fairness in Recommender Systems}
The recommender systems research community has begun to consider issues of fairness in recommendation. A frequently practiced strategy for encouraging fairness is to enforce \emph{demographic parity} among different protected groups.
Demographic parity aims to ensure that the set of individuals in each protected group have similar overall distributions over outcomes (e.g. recommended items) ~\cite{zemel2013learning}. Some authors have addressed the unfairness issue in recommender systems by adding a regularization term that enforces demographic parity~\cite{kamishima2011fairness,kamishima2012enhancement,kamishima2013efficiency,kamishima2014correcting,kamishima2016model}.  
However, demographic parity is only appropriate when user preferences have no legitimate relationship to the protected attributes. In recommendation systems, user preferences are indeed often influenced by protected attributes such as gender, race, and age~\cite{chausson2010watches}. Therefore, enforcing demographic parity may significantly damage the quality of recommendations. Fair recommendation systems have also been proposed by penalizing disparate distributions of prediction error~\cite{yao2017beyond}, and by making recommended items independent from protected attributes such as gender, race, or age~\cite{kamishima2017considerations}.  In addition, ~\cite{burke2017balanced,burke2017multisided} taxonomize fairness objectives and methods based on which set of stakeholders in the recommender system are being considered, since it may be meaningful to consider fairness among many different groups. Pareto efficiency-based fairness-aware group recommendation~\cite{xiao2017fairness} was also proposed, however this method is not effective in personalized fair recommendations. 

In a non-archival extended abstract~\cite{rashid}, we recently proposed a simple technique to improve fairness in social media-based CF models.  Our approach was to learn an NCF model, then debias the user embeddings using a linear projection technique, and predict sensitive items using k-nearest neighbors or logistic regression. This method improves the fairness of CF, but substantially degrades the accuracy of recommendations. In this work, we 
improve on this approach by using the method in ~\cite{rashid} to debias a pre-trained neural model for non-sensitive items, then fine-tuning using a fairness penalty to learn to recommend sensitive items. Our results show improved fairness and accuracy versus~\cite{rashid}.

\subsection{Fairness Metrics}
We consider several existing fairness metrics which are applicable for collaborative filtering problems.
\subsubsection{Differential Fairness}
The differential fairness \cite{foulds2018intersectional,foulds2018bayesian} metric aims to ensure equitable treatment for all protected groups, and it provides a privacy interpretation of disparity.  Let $M(x)$ be an algorithmic mechanism (e.g. a recommender system) which takes an individual's data $x$ and assigns them an outcome $y$ (e.g. a class label or whether a user-item interaction is present). The mechanism $M(x)$ is $\epsilon$-\emph{differentially fair (DF)} with respect to $(A, \Theta)$ if for all $\theta \in \Theta$ with $x \sim \theta$, and $y \in \mbox{Range}(M)$,
\begin{equation}
	e^{-\epsilon} \leq \frac{P_{M, \theta}(M(x) = y|s_i, \theta)}{P_{M, \theta}(M(x) = y|s_j, \theta)}\leq e^\epsilon \mbox{ ,} 
\end{equation}
for all $(s_i, s_j) \in A \times A$ where $P(s_i|\theta) > 0$, $P(s_j|\theta) > 0$. Here, $s_i$, $s_j \in A$ are tuples of all protected attribute values, e.g. male  and female, and $\Theta$, the set of data generating distributions, is typically a point estimate of the data distribution. If all of the $P_{M, \theta}(M(x) = y|s, \theta)$ probabilities are equal for each group $s$, across all outcomes $y$ and distributions $\theta$, $\epsilon = 0$, otherwise $\epsilon > 0$. 
\cite{foulds2018intersectional} proved that a small $\epsilon$ guarantees similar utility per protected group, and ensures that protected attributes cannot be inferred based on outcomes. 
DF can be estimated using smoothed ``soft counts'' of the predicted outcomes based on a probabilistic model.  For gender bias in our recommender (assuming a gender binary), we can estimate $\epsilon$-DF per sensitive item $i$ by verifying that:
\begin{align}
e^{-\epsilon} 
\leq \frac{\sum_{u: A = m} \hat{y}_{ui} + \alpha}{N_{m}  + 2\alpha}\frac{N_{f} + 2\alpha}{\sum_{u: A = f} \hat{y}_{ui} + \alpha}\leq e^\epsilon \label{eqn:smoothedFairnessSoft}  \mbox{ ,}\nonumber  \\ 
e^{-\epsilon} 
\leq \frac{\sum_{u: A = m} (1-\hat{y}_{ui}) + \alpha}{N_{m}  + 2\alpha}\frac{N_{f} + 2\alpha}{\sum_{u: A = f} (1-\hat{y}_{ui}) + \alpha}\leq e^\epsilon 
\mbox{ ,} 
\end{align} 
where scalar $\alpha$ is each entry of the parameter of a symmetric Dirichlet prior with concentration parameter $2\alpha$, $i$ is an item and $N_{A}$ is the number of users of gender $A$ ($m$ or $f$).

\subsubsection{Absolute Unfairness}
The absolute unfairness ($U_{abs}$) metric for recommender systems measures the discrepancy between the predicted behavior for disadvantaged and advantaged users~\cite{yao2017beyond}. It measures inconsistency in absolute estimation error across user types, defined as follows:
\begin{equation}
    U_{abs} = \frac{1}{N}\sum_{j=1}^{N}||(E_{D}[\hat{y}_{ui}]_j-E_{D}[r]_j)|-|(E_{A}[\hat{y}_{ui}]_j-E_{A}[r]_j)||
\end{equation}
where, for $N$ items, $E_{D}[\hat{y}_{ui}]_j$ is the average predicted score for the $j$-th item for disadvantaged users, $E_{A}[\hat{y}_{ui}]_j$ is the average predicted score for advantaged users, and$E_{D}[r]_j$ and $E_{A}[r]_j$ are the average score for the disadvantaged and advantaged users, respectively. $U_{abs}$ captures a single statistic representing the quality of prediction for each user group. If one protected group has small estimation error and the other group has large estimation error, then the former type of group has the unfair advantage of good recommendations, while the other user group has poor recommendations.

\section{Neural Fair Collaborative Filtering}
\begin{figure*}[t]
		\centerline{\includegraphics[width=0.85\textwidth]{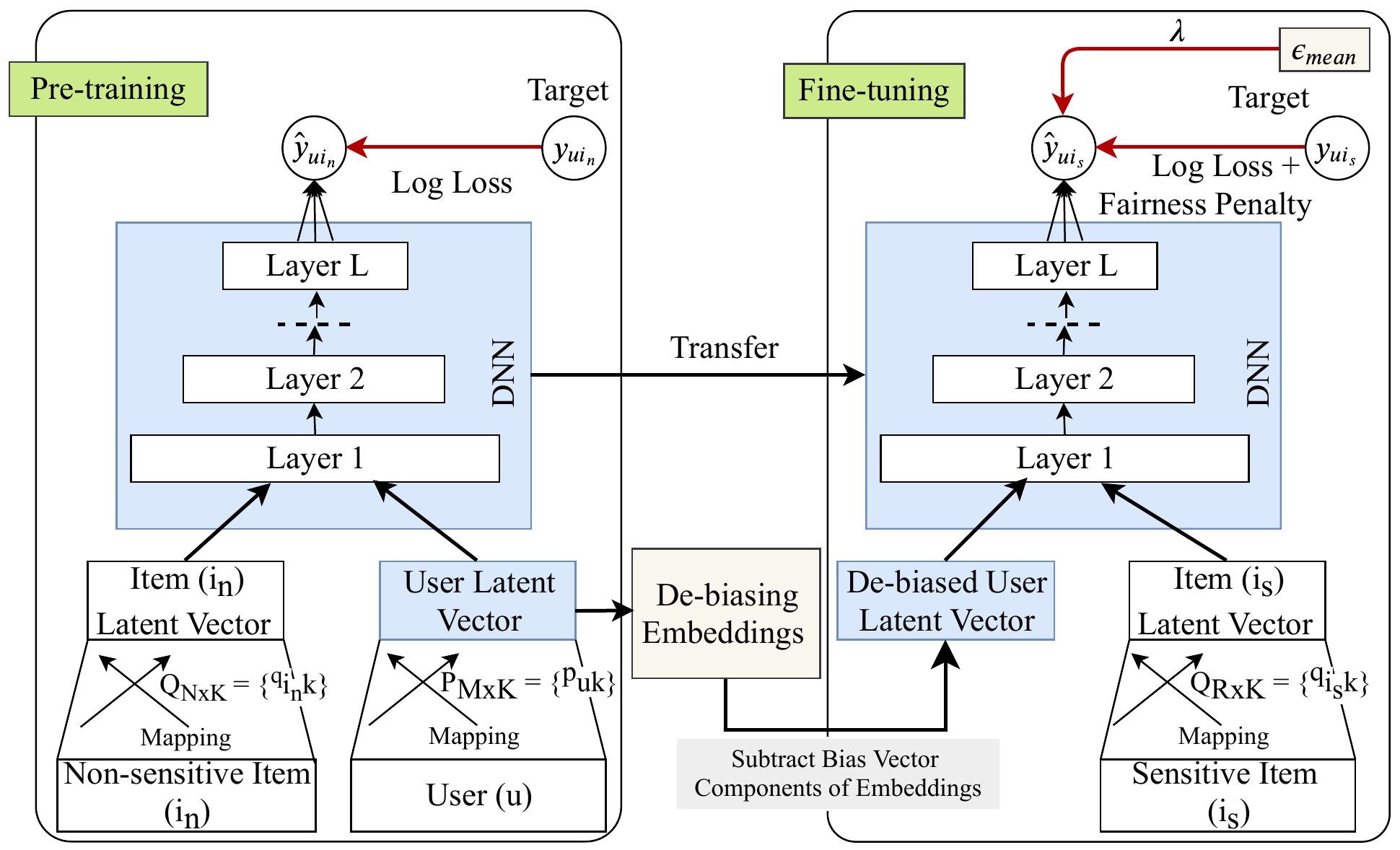}}
		\caption{\small Schematic diagram of neural fair collaborative filtering (NFCF). Red arrows indicate back-propagation only.  
		}
		\label{fig:NFCF_block}
\end{figure*}

Due to biased data that encode harmful human stereotypes in our society, typical social media-based collaborative filtering (CF) models can encode gender bias and make unfair decisions. In this section, we propose a practical framework to mitigate gender bias in CF recommendations, which we refer to as \emph{neural fair collaborative filtering} (NFCF) as shown in Figure~\ref{fig:NFCF_block}. The main components in our NFCF framework are as follows: an \emph{NCF model}, \emph{pre-training user and non-sensitive item embeddings}, \emph{de-biasing pre-trained user embeddings}, and \emph{fine-tuning with a fairness penalty}.
We use NCF as the CF model because of its flexible network structure for pre-training and fine-tuning. We will show the value of each component below with an \emph{\textbf{ablation study}} (Table \ref{tab:ablation_study}). Similarly to \cite{he2017neural}, the DNN under the NCF model can be defined as:
\begin{align}\label{eq:NCF}
    {z}_1 = \phi_1({p}_u,{q}_i) = {\begin{bmatrix}{p}_u\\{q}_i \end{bmatrix}} \mbox{ , } \nonumber\\
    {z}_2 = \phi_2(z_1) = a_2(W_{2}^{T}z_1 + b_2) \mbox{ , } \nonumber\\
    \vdots \nonumber\\
    \phi_L(z_{L-1}) = a_L(W_{L}^{T}z_{L-1} + b_L) \mbox{ , } \nonumber\\
    \hat{y}_{ui} = \sigma(h^{T}\phi_L(z_{L-1}))
\end{align}
where $z_l$, $\phi_l$, $W_l$, $b_l$. and $a_l$ denote the neuron values, mapping function
, weight matrix, intercept term, and activation function for the $l$-th layer's perceptron, respectively. The DNN is applied to ${z}_1$ to learn the user-item latent interactions.
\begin{algorithm}[t]
\caption{Training NFCF for Gender De-biased Recommendations}\label{alg-nfcf}
\small
\begin{flushleft}
\textbf{Input:} user and non-sensitive item pairs: $\mathcal{D}_n = (u,i_{n})$, user and sensitive item pairs: $\mathcal{D}_s = (u,i_{s})$, and gender attribute: $A$\\
\textbf{Output:} Fair CF model $M_\mathbf{W}$(x) for $i_{s}$ recommendations \\
\end{flushleft}

\begin{flushleft}
\emph{\textbf{Pre-training steps:}}
\end{flushleft}
\begin{itemize}
\item Randomly initialize $M_\mathbf{W}$(x)'s parameters $\boldsymbol{W}$: $p_u$, $q_{i_n}$, $W_l$, and $b_l$
\item For each epoch of ${D}_n$ :
\begin{itemize}
    \item For each mini-batch:
    \begin{itemize}
        \item Learn $M_\mathbf{W}$(x)'s parameters $\boldsymbol{W}$ by minimizing:
        \item[] $L = -\sum_{(u,i_n)\in \chi \cup \chi^{-}}[ y_{ui_{n}}\log\hat{y}_{ui_{n}} + (1-y_{ui_{n}})\log(1-\hat{y}_{ui_{n}})]$
    \end{itemize}
\end{itemize}
\end{itemize}
\begin{flushleft}
\emph{\textbf{De-biasing embeddings steps:}}
\end{flushleft}
\begin{itemize}
    \item Compute gender bias vector $v_B$ using Equation ~\ref{eq:female} and ~\ref{eq:bias}
    \item De-bias each user embedding using: $p_{u} := p_{u} - (p_{u}\cdot v_{B})v_{B}$
\end{itemize}
\begin{flushleft}
\emph{\textbf{Fine-tuning steps:}}
\end{flushleft}
\begin{itemize}
\item Initialize with pre-trained $M_\mathbf{W}$(x)'s parameters $\boldsymbol{W}$: $W_l$, $b_l$ and de-biased $p_{u}$ with randomly initialized $q_{i_s}$  
\item For each epoch of ${D}_s$ :
\begin{itemize}
    \item For each mini-batch:
    \begin{itemize}
        \item Fine-tune  $M_\mathbf{W}$(x) by minimizing (while $p_{u}$ is kept fixed):
        \item[] \quad\quad $\underset{\textbf{W}}{\text{min}}[L_{\mathbf{\chi \cup \chi^{-}}}(\textbf{W}) + \lambda R_{\mathbf{\chi}}(\epsilon_{mean})]$
    \end{itemize}
\end{itemize}
\end{itemize}
\end{algorithm}

In the first step of our NFCF method, \emph{pre-training user and item embeddings}, NCF is trained to predict users' interactions with \emph{non-sensitive} items (e.g. ``liked'' social media pages) via back-propagation.  This leverages plentiful non-sensitive social media data to learn user embeddings and network weights, but may introduce \textbf{demographic bias due to correlations between non-sensitive items and demographics}.  E.g., liking the \emph{Barbie doll} page typically correlates with user gender.  These correlations are expected to result in systematic differences in the embeddings for different demographics, which in turn can lead to systematic differences in sensitive item recommendations. To address this, in step two the user embeddings from step one are \emph{de-biased}. Our method to de-bias user embeddings adapts  a very recent work on attenuating bias in word vectors~\cite{dev2019attenuating} to the task of collaborative filtering. Specifically, ~\cite{dev2019attenuating} propose to debias word vectors using a linear projection of each word embedding $w$ orthogonally onto a \emph{bias vector} $v_{B}$, which identifies the ``bias component'' of $w$.  The bias component is then removed via $w' = w - (w\cdot v_{B})v_B$. 

To adapt this method to CF, the main challenge is to find the proper bias direction $v_B$. ~\cite{dev2019attenuating} construct $v_B$ based on word embeddings for gender-specific first names, which are not applicable for CF.  
We instead use \emph{CF embeddings for users from each protected group}.  
We first compute a group-specific bias direction for female users as
\begin{equation}\label{eq:female}
    v_{female} = \frac{1}{n_f}(f_1+f_2+\dots +f_n)\mbox{ ,}
\end{equation}
where $f_1,f_2,\dots$ are vectors for each female user, and  $n_f$ is the total number of female users.  We similarly compute a bias direction for men $v_{male}$. Finally, we compute the overall gender bias vector:
\begin{equation}\label{eq:bias}
    v_{B} = \frac{v_{female}-v_{male}}{||v_{female}-v_{male}||}\mbox{ .}
\end{equation}
We then de-bias each user embedding $p_{u}$ by subtracting its component in the direction of the bias vector:
\begin{equation}
    p_{u}' = p_{u} - (p_{u}\cdot v_{B})v_{B} \mbox{ .}
\end{equation}
As we typically do not have demographic attributes for items, we only de-bias user embeddings and not item embeddings.   
In the third step, we \emph{transfer} the de-biased user embeddings and pre-trained DNN's parameters to a model for recommending \emph{sensitive items}, which we \emph{fine-tune} for this task.  
During fine-tuning, a \emph{fairness penalty} is added to the objective function to address a second source of bias: \textbf{demographic bias in the sensitive items}.  E.g., more men than women choose computer science  careers~\cite{broad2014recruiting}, and this should be corrected~\cite{beede2011women}.   
We penalize the mean 
of the \emph{per-item} $\epsilon$'s:
\begin{equation}
    \epsilon_{mean} = \frac{1}{n_s}\sum_{i=1}^{n_s}\epsilon_i  \mbox{ ,} 
\end{equation}
where $\epsilon_1, \epsilon_2, \dots \epsilon_{n_s}$ are the DF measures for sensitive items and $\epsilon_{mean}$ 
is the average across the $\epsilon$'s for each item. 
Following ~\cite{foulds2018intersectional}, our learning algorithm for fine-tuning uses the fairness cost as a regularizer to balance the trade-off between fairness and accuracy. Using back-propagation, we minimize the loss function $L_{\mathbf{\chi \cup \chi^{-}}}(\textbf{W})$ from Equation~\ref{eq:loss} for model parameters $\textbf{W}$ plus a penalty on $\epsilon_{mean}$, weighted by a tuning parameter $\lambda>0$:
\begin{equation}\label{eq:objective}
    \underset{\textbf{W}}{\text{min}}[L_{\mathbf{\chi \cup \chi^{-}}}(\textbf{W}) + \lambda R_{\mathbf{\chi}}(\epsilon_{mean})]
\end{equation}
where $R_{\mathbf{\chi}}(\epsilon_{mean}) =max(0,\epsilon_{mean_{M_\mathbf{W}(\mathbf{\chi})}} - \epsilon_{mean_{0}})$ is the fairness penalty term, and $\epsilon_{mean_{M_\mathbf{W}(\mathbf{\chi})}}$ is the $\epsilon_{mean}$ for the CF model $M_\mathbf{W}(\mathbf{\chi})$ while $\chi$ and $\chi^{-}$ are the set of interacted and not-interacted user-item pairs, respectively.
Setting the target $\epsilon_{mean_{0}}$ to 0 encourages demographic parity, while setting $\epsilon_{mean_{0}}$ to the dataset's empirical value penalizes any increase in the unfairness metric over ``societal bias'' in the data (which would in this case be presumed to be legitimate).  In our experiments, we use $\epsilon_{mean_{0}}=0$. Pseudo-code for training the \emph{NFCF} algorithm is given in Algorithm~\ref{alg-nfcf}.

We also consider an additional variant of the proposed framework, called \emph{NFCF\_embd}, which only  de-biases  the user embeddings to mitigate bias. In this algorithm, we compute the bias vector $v_B$ on the pre-trained user embeddings, fine-tune the model for sensitive item recommendations without any fairness penalty, and then de-bias the held-out user embeddings using the pre-computed bias vector. Since there is no additional fairness penalty with the objective, this algorithm converges faster. There is also no requirement to tune the $\lambda$ hyperparameter.
\section{Experiments}
In this section, we 
validate and compare our model with multiple baselines for recommending careers and academic concentrations using social media data. 
Our implementation's source code will be provided in the online. 
\subsection{Datasets}
\begin{table}[t]
	\centering
	\small
	\resizebox{1.0\textwidth}{!}{
    \begin{tabular}{llllllllllll}
    \toprule
      & \multicolumn{4}{c}{Non-sensitive Data} &  \multicolumn{6}{c}{Sensitive Data} \\
      \midrule
      & Users  & Items   & Pairs   & Sparsity & & Users & Males  & Females  & Items & Pairs    & Sparsity \\
      \cline{2-5} \cline{7-12}
      \emph{MovieLens} Dataset & 6,040  & 3,416    & 999,611 & 95.16\%  & &  4,920 & 3,558  & 1,362    & 17      & 4,920    & 94.12\% \\
      \emph{Facebook} Dataset & 29,081 & 42,169  & 5,389,541 & 99.56\%  & & 13,362 & 5,053  & 8,309   & 169       & 13,362   & 99.41\%         \\
    \bottomrule
    \end{tabular}
    }
	\caption{Statistics of the datasets.
	\label{tab:data}}
\end{table}
\begin{figure*}[t]
		\centerline{\includegraphics[width=0.98\textwidth]{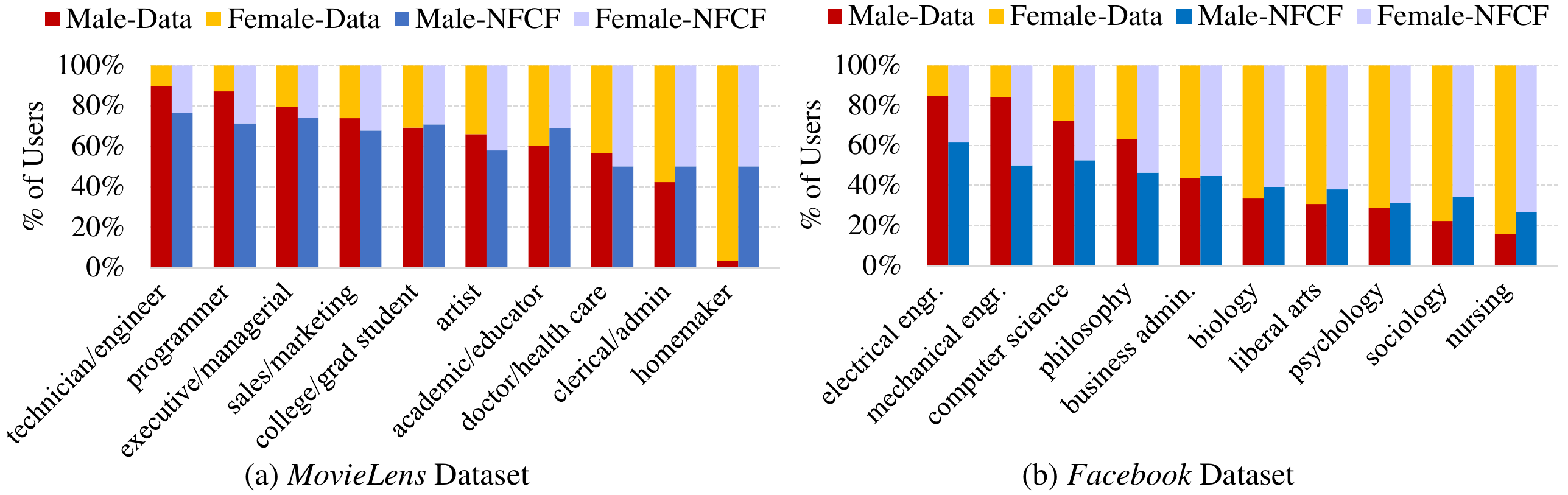}}
		\caption{\small Gender distributions of example gender-biased careers and college majors for (a) \emph{MovieLens} and (b) \emph{Facebook} datasets.  We report the distributions in the dataset (left columns), and corresponding top-$1$ recommendation by our NFCF model (right columns).
		}
		\label{fig:user_dist_all}
\end{figure*}
We evaluate our models on two datasets: \emph{MovieLens},\footnote{\url{http://grouplens.org/datasets/movielens/1m/}.} a public dataset which facilitates research reproducibility, and a \emph{Facebook} dataset which is larger and is a more realistic setting for a fair social media-based recommender system. 
\subsubsection{\textbf{MovieLens Data}}
We analyzed the widely-used \emph{MovieLens} dataset which contains 1 million ratings of 3,900 movies by 6,040 users who joined MovieLens~\cite{harper2015movielens}, a noncommercial movie recommendation service operated by the University of Minnesota.
We used \emph{gender} as the protected attribute, self-reported \emph{occupation} as the sensitive item (with one occupation per user), and \emph{movies} as the non-sensitive items.  
Since we focus on implicit feedback, which is common in a social media setting (e.g. page ``likes''), we converted explicit movie ratings to binary implicit feedback~\cite{koren2008factorization,he2017neural}, where a 1 indicates that the user has rated the item. We discarded movies that were rated less than 5 times, and users who declared their occupation as ``K-12 student,'' ``retired,'' ``unemployed,'' and ``unknown or not specified'' were discarded for career recommendation.  A summary of the pre-processed dataset is shown in Table~\ref{tab:data}. 
\subsubsection{\textbf{Facebook Data}}
The \emph{Facebook} dataset we analyzed was collected as part of the myPersonality project ~\cite{kosinski2015facebook}. The data for research were collected with opt-in consent.  We used \emph{gender} as the protected attribute,  \emph{college major} as the sensitive items (at most one per user), and \emph{user-page} interaction pairs as the non-sensitive items. A user-page interaction occurs when a user ``likes'' a Facebook page.  We discarded pages that occurred in less than 5 user-page interactions.  See Table~\ref{tab:data} for a summary of the dataset after pre-processing. 

\subsubsection{\textbf{Gender Distributions for Datasets}}
In Figure~\ref{fig:user_dist_all}, we show  disparities in the gender distributions of $10$ example careers and college majors for \emph{MovieLens} and \emph{Facebook} datasets, respectively. For example, 97\% of the associated users for the occupation \emph{homemaker} are women in the \emph{MovieLens} data, while there are only 27\% women among the users associated with the \emph{computer science} major in the \emph{Facebook} data. As a qualitative illustration, we also show the gender distribution of top-1 recommendations from our proposed NFCF model. NFCF mitigated gender bias for most of these sensitive items. In the above examples, NFCF decreased the percentage of women for \emph{homemaker} from 97\% to 50\%, while increasing the percentage of women for \emph{computer science} from 27\%  to 48\%.      
\subsection{Baselines}
\begin{table}[t]
	\centering
	\small
    \begin{tabular}{lcccc}
    \hline
    \multicolumn{5}{c}{\emph{MovieLens} Dataset}\\
    \hline
    Models & HR@$10$        & NDCG@$10$      & HR@$25$        & NDCG@$25$      \\ \hline
    NCF    & 0.543 & \textbf{0.306} & 0.825 & \textbf{0.377} \\
    MF     & \textbf{0.551}          & 0.304          & \textbf{0.832}          & 0.374          \\ \hline
    \end{tabular}
    \hspace{0.25 cm}
	\small
    \begin{tabular}{lcccc}
    \hline
    \multicolumn{5}{c}{\emph{Facebook} Dataset}\\
    \hline
    Models & HR@$10$        & NDCG@$10$      & HR@$25$        & NDCG@$25$      \\ \hline
    NCF    & \textbf{0.720} & \textbf{0.468} & \textbf{0.904} & \textbf{0.514} \\
    MF     & 0.609          & 0.382          & 0.812          & 0.434          \\ \hline
    \end{tabular}
	\caption{Performance of NCF and MF models for movie and Facebook page recommendations (the pre-training task) on the \emph{Movielens} and \emph{Facebook} datasets, respectively. 
	\label{tab:preTrain_performance}}
\end{table}
%
\begin{figure*}[t]
		\centerline{\includegraphics[width=0.9\textwidth]{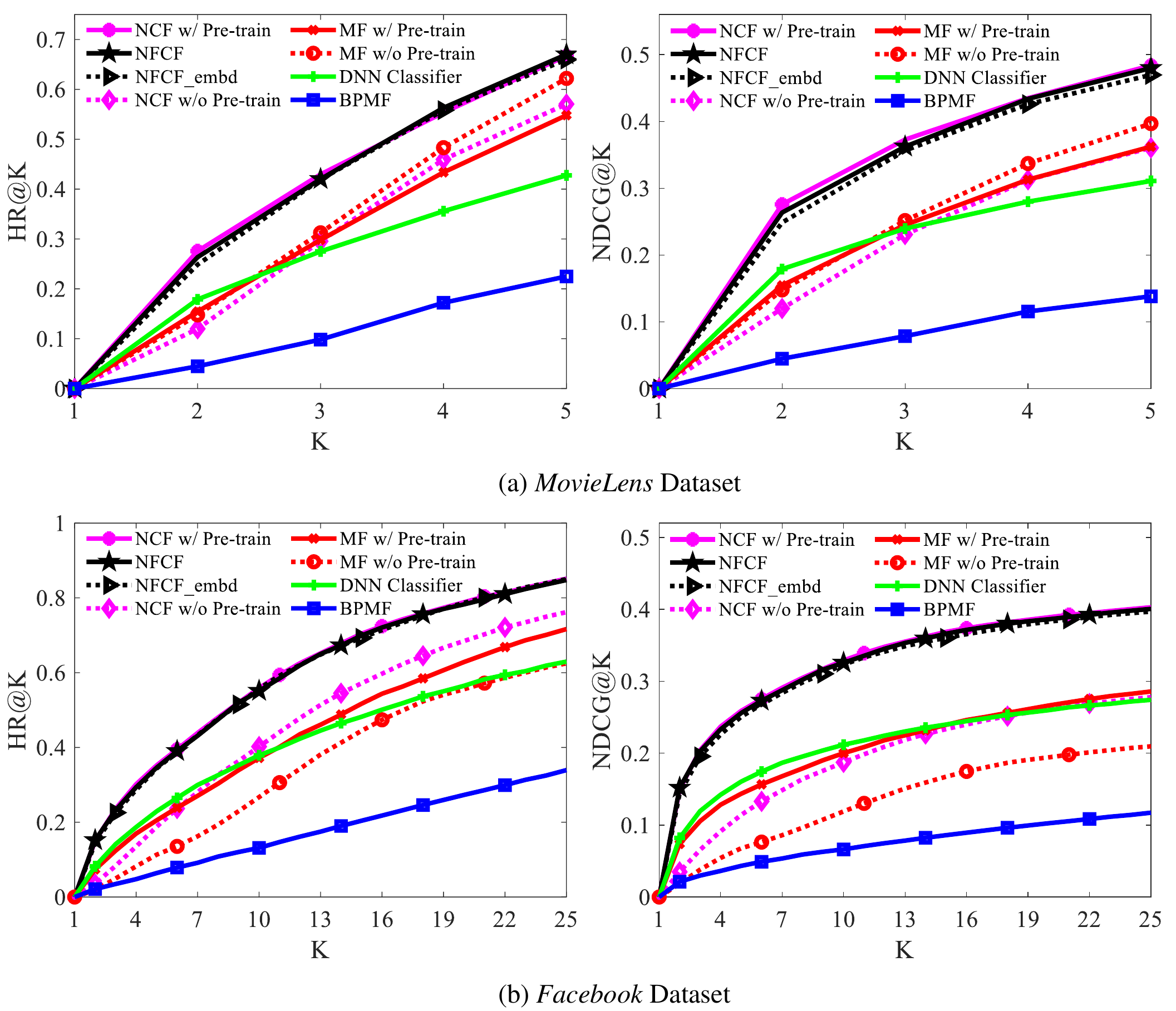}}
		\caption{\small Comparison of proposed models with ``typical'' baselines that do not consider fairness. Evaluation of Top-$K$ career and college major recommendations on the (a) \emph{MovieLens} (among 17 unique careers) and (b) \emph{Facebook} (among 169 unique majors) datasets, where $K$ ranges from $1$ to $5$ and $1$ to $25$, respectively. \emph{NCF w/ Pre-train} outperforms all the baselines; NFCF performs similarly.}
		\label{fig:preTrain}
\end{figure*}
\begin{figure*}[t]
		\centerline{\includegraphics[width=0.95\textwidth]{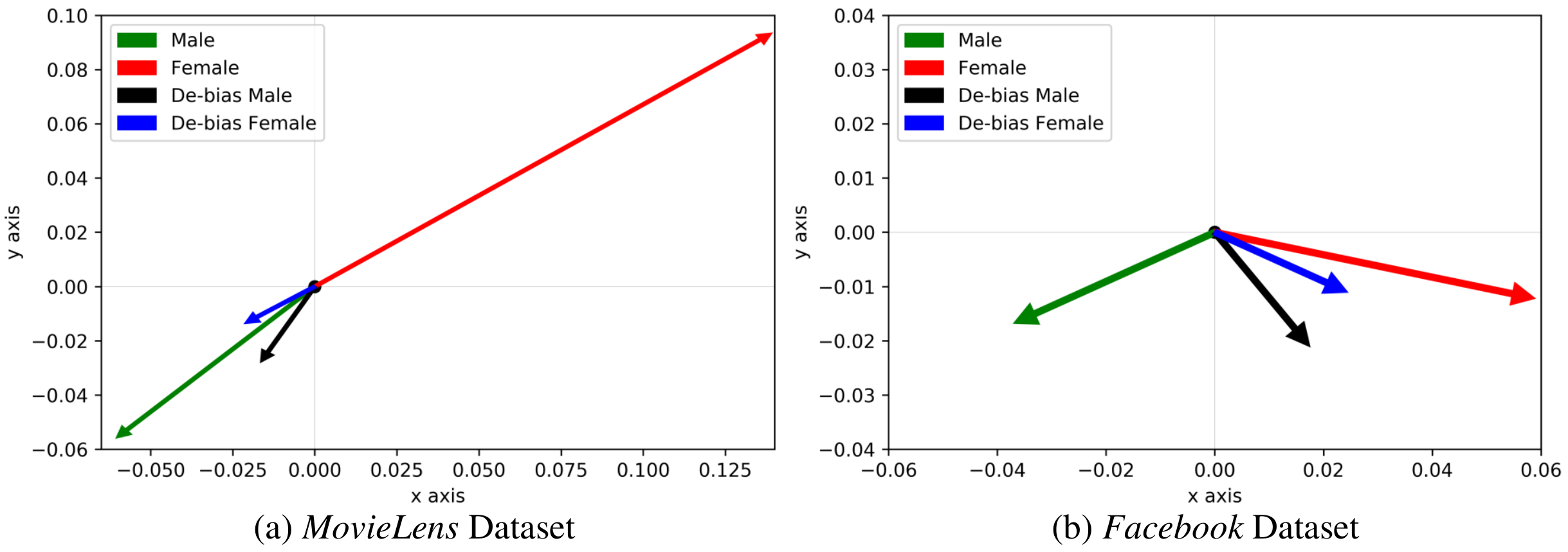}}
		\caption{\small De-biasing pre-trained user embeddings by 
		removing the component along the bias direction $v_B$ (PCA projection) for the (a) \emph{MovieLens} and (b) \emph{Facebook} datasets. PCA was performed based on all embeddings.}
		\label{fig:debiasing_vectors}
\end{figure*}
We compare our proposed framework to the following ``typical'' baseline models without any fairness constraints:
\begin{itemize}
    \item \textbf{MF w/o Pre-train}. Typical matrix factorization (MF) model which is trained with the user-item interactions for sensitive item recommendations, where the items contain both non-sensitive and sensitive items.
    \item \textbf{MF w Pre-train}. Typical MF model which is pre-trained with the interactions of user and non-sensitive items and fine-tuned with the interactions of users and sensitive items. Specifically, $q_{i}$, $b_i$, and $b_u$ from Equation~\ref{eq:MF} are fine-tuned while $p_{u}$ is kept fixed.
    \item \textbf{NCF w/o Pre-train}. Typical NCF model which is trained with the user-item interactions for sensitive item recommendations, where the items contain both non-sensitive and sensitive items.
    \item \textbf{NCF w Pre-train}. Typical NCF model which is pre-trained with the interactions of users and non-sensitive items and fine-tuned with the interactions of users and sensitive items.  Specifically, $q_{i}$, ${W}_{l}$, and ${b}_{l}$ from Equation~\ref{eq:NCF} are fine-tuned while $p_{u}$ is kept fixed.
    \item \textbf{DNN Classifier}. A simple baseline where we train a DNN-based classifier to predict career labels given the interactions of users and non-sensitive items as features (i.e. binary features, one per user-page ``like'' or one per user-movie ``rating''). No user embeddings are learned. 
    \item \textbf{BPMF}. We also used Bayesian probabilistic matrix factorization (BPMF) via MCMC~\cite{salakhutdinov2008bayesian} as a baseline, since it typically has good performance with small data. BPMF is trained on the user-item interactions for sensitive item recommendations, where the items contain both non-sensitive and sensitive items.
\end{itemize}
We also compared our proposed models with the following fair baseline models: 
\begin{itemize}
    \item \textbf{Projection-based CF}. This is our previous linear projection-based fair CF method~\cite{rashid}. First, NCF is trained using user and non-sensitive item data, followed by de-biasing user vectors 
    by subtracting the component in the bias direction. 
    Finally, a multi-class logistic regression model is trained on the de-biased user vectors to predict sensitive items.  
    \item \textbf{MF-$U_{abs}$}. The learning objective of the MF model is augmented with a smoothed variation of $U_{abs}$~\cite{yao2017beyond} using the Huber loss~\cite{huber1992robust}, weighted by a tuning parameter $\lambda$. The MF-$U_{abs}$ model is trained with the user-item interactions for career recommendations, where the items include both non-sensitive and sensitive items.
    \item \textbf{Resampling for Balance}. This method~\cite{ekstrand2018all} involves pre-processing by resampling the training user-item data to produce a gender-balanced version of the data. First, we extract user-item data for users with known gender information and randomly sample the same number of male and female users without replacement. The items include both non-sensitive and sensitive items. Finally, NCF and MF are trained on the gender-balanced user-item data to perform sensitive item recommendation. 
\end{itemize}
\begin{table*}[t]
	\centering
	\small
    \begin{tabular}{lcccccc}
    \toprule
    \multicolumn{7}{c}{\emph{MovieLens} Dataset}\\
    \midrule
    Ablation study                     & HR@$5\uparrow$ & NDCG@$5\uparrow$ & HR@$7\uparrow$ & NDCG@$7\uparrow$ & $\epsilon_{mean}\downarrow$ & $U_{abs}\downarrow$ \\ \midrule
    NFCF                      & \textbf{0.670}  & 0.480    & 0.822  & 0.536    & \textbf{0.083}             & \textbf{0.009}     \\
    w/o pre-train             & 0.493  & 0.323    & 0.731  & 0.446    & 0.112             & 0.017     \\
    w/o de-biasing embeddings & 0.665  & \textbf{0.481}    & \textbf{0.832}  & \textbf{0.543}    & 0.120             & 0.010     \\
    w/o fairness penalty      & 0.667  & 0.480    & 0.827  & 0.539    & 0.097             & 0.013     \\
    replace NCF w/ MF         & 0.514  & 0.350    & 0.707  & 0.423    & 0.122             & 0.021     \\ \bottomrule
    &&&&&&\\
    \toprule
    \multicolumn{7}{c}{\emph{Facebook} Dataset}\\
    \midrule
    Ablation study            & HR@$10\uparrow$ & NDCG@$10\uparrow$ & HR@$25\uparrow$ & NDCG@$25\uparrow$ & $\epsilon_{mean}\downarrow$ & $U_{abs}\downarrow$ \\ \midrule
    NFCF                      & 0.551   & 0.326     & 0.848   & \textbf{0.401}       & \textbf{0.302}             & \textbf{0.024}     \\
    w/o pre-train             & 0.339   & 0.127     & 0.587   & 0.224       & 0.613             & 0.038     \\
    w/o de-biasing embeddings & 0.556   & \textbf{0.328}     & 0.847   & 0.400       & 0.314             & \textbf{0.024}     \\
    w/o fairness penalty      & \textbf{0.557}   & 0.327     & \textbf{0.849}   & \textbf{0.401}       & 0.363             & 0.026     \\
    replace NCF w/ MF         & 0.297   & 0.112     & 0.427   & 0.194       & 0.880             & 0.071     \\ \bottomrule
    \end{tabular}
	\caption{Ablation study of \emph{NFCF} 
	for career and college major recommendations on the \emph{MovieLens} and \emph{Facebook} datasets. Higher is better for HR and NDCG; lower is better for $\epsilon_{mean}$ and $U_{abs}$. Removing each model component harms performance and/or fairness.
	\label{tab:ablation_study}}
\end{table*}
\begin{figure*}[t]
		\centerline{\includegraphics[width=0.90\textwidth]{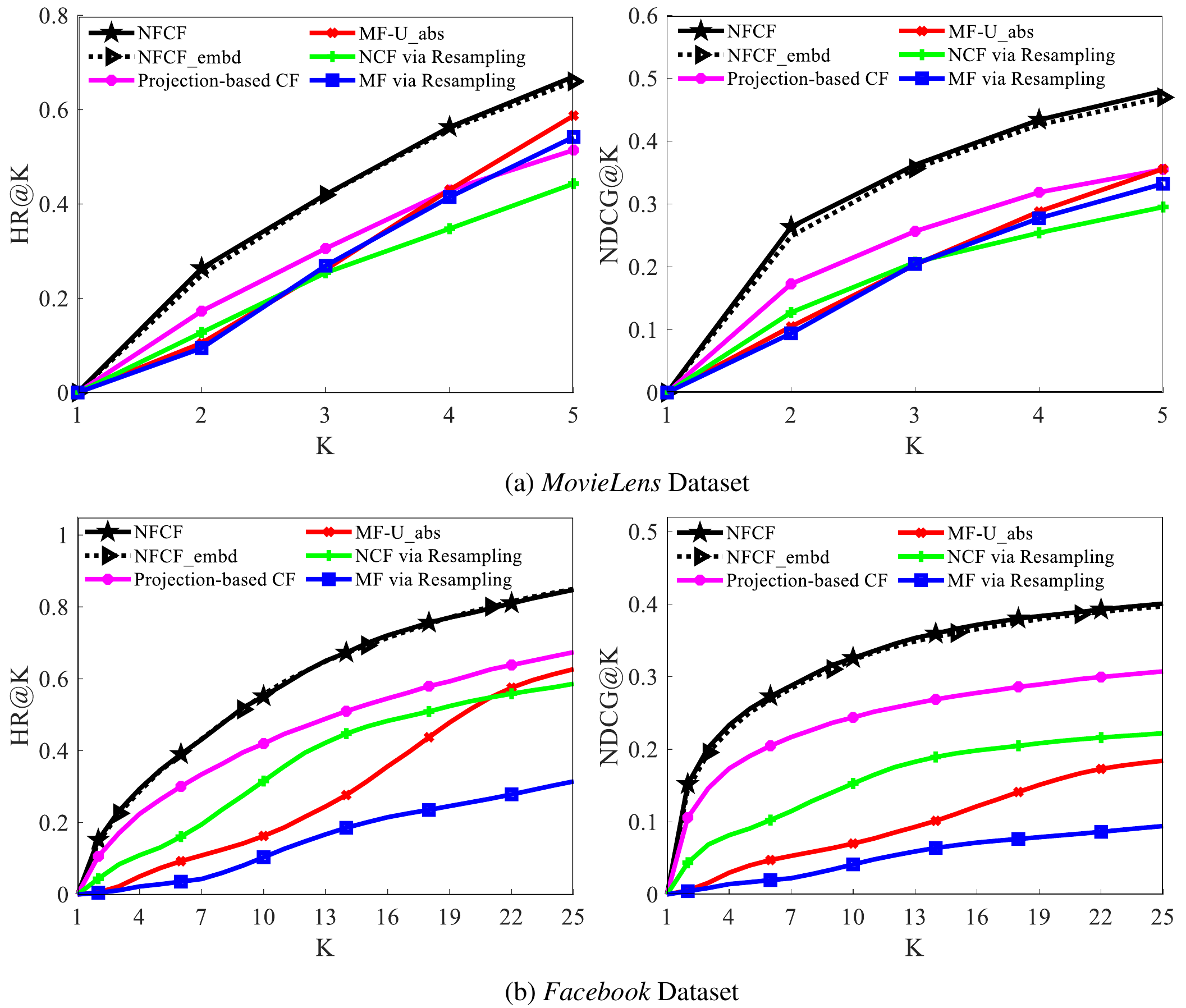}}
		\caption{\small Comparison of proposed models with fair baselines. Evaluation of Top-$K$ career and college major recommendations on the (a) \emph{MovieLens} (among 17 unique careers) and (b) \emph{Facebook} (among 169 unique majors) datasets, where $K$ ranges from $1$ to $5$ and $1$ to $25$, respectively. \emph{NFCF} outperforms all the baselines; \emph{NFCF\_embd} performs similarly.}
		\label{fig:compare_fair_model}
\end{figure*}
%
\begin{table*}[t]
	\centering
    \small
    \resizebox{1.0\textwidth}{!}{
    \begin{tabular}{llcccccc}
    \toprule
    \multicolumn{8}{c}{\emph{MovieLens} Dataset}\\
    \midrule
    & Models            & HR@$5\uparrow$         & NDCG@$5\uparrow$       & HR@$7\uparrow$         & NDCG@$7\uparrow$       & $\epsilon_{mean}\downarrow$ & $U_{abs}\downarrow$      \\ \midrule
    \multirow{2}{*}{Proposed Models} & NFCF              & \textbf{0.670} & 0.480          & 0.822          & 0.536          & \textbf{0.083}    & \textbf{0.009} \\
    & NFCF\_embd        & 0.661          & 0.470          & \textbf{0.825} & 0.531          & 0.091             & 0.016          \\
    \midrule
    \multirow{5}{*}{Typical Baselines} & NCF w Pre-train   & 0.667          & \textbf{0.484} & \textbf{0.825} & \textbf{0.542} & 0.188             & 0.022          \\
    & NCF w/o Pre-train & 0.570          & 0.360          & 0.762          & 0.432          & 0.244             & 0.026          \\
    & MF w Pre-train    & 0.548          & 0.362          & 0.747          & 0.436          & 0.285             & 0.060          \\ 
    & MF w/o Pre-train  & 0.622          & 0.397          & 0.820          & 0.471          & 0.130             & 0.020          \\
    & DNN Classifier    & 0.428          & 0.311          & 0.546          & 0.355          & 0.453             & 0.035          \\ 
    & BPMF    & 0.225          & 0.138          & 0.338          & 0.180          & 0.852             & 0.063          \\ 
    \midrule
    \multirow{4}{*}{Fair Baselines} & Projection-based CF ~\cite{rashid} & 0.514          & 0.355          & 0.655          & 0.408          & 0.229             & 0.012          \\
    & MF-$U_{abs}$ ~\cite{yao2017beyond}     & 0.588          & 0.356          & 0.776          & 0.426          & 0.096             & 0.017          \\
    & NCF via Resampling ~\cite{ekstrand2018all} & 0.443         & 0.295          & 0.622          & 0.362          & 0.144             & 0.022          \\
    & MF via Resampling ~\cite{ekstrand2018all} & 0.542          & 0.332          & 0.759          & 0.413          & 0.103             & 0.029          \\
    \bottomrule
    &&&&&&&\\
    \toprule
    \multicolumn{8}{c}{\emph{Facebook} Dataset}\\
    \midrule
    & Models            & HR@$10\uparrow$        & NDCG@$10\uparrow$      & HR@$25\uparrow$        & NDCG@$25\uparrow$      & $\epsilon_{mean}\downarrow$ & $U_{abs}\downarrow$      \\ \midrule
    \multirow{2}{*}{Proposed Models} & NFCF              & 0.551          & 0.326          & 0.848          & 0.401          & \textbf{0.302}    & 0.024          \\
    & NFCF\_embd           & 0.557          & \textbf{0.333} & 0.850          & 0.397          & 0.359             & \textbf{0.022} \\
    \midrule
    \multirow{5}{*}{Typical Baselines} & NCF w Pre-train   & \textbf{0.559} & 0.329          & \textbf{0.851} & \textbf{0.403} & 0.376             & 0.027          \\
    & NCF w/o Pre-train & 0.402          & 0.187          & 0.762          & 0.278          & 0.785             & 0.039          \\
    & MF w Pre-train    & 0.372          & 0.200          & 0.717          & 0.286          & 0.875             & 0.077          \\ 
    & MF w/o Pre-train  & 0.267          & 0.119          & 0.625          & 0.210          & 0.661             & 0.029          \\
    & DNN Classifier    & 0.379          & 0.212          & 0.630          & 0.274          & 0.633             & 0.070          \\ 
    & BPMF              & 0.131          & 0.066          & 0.339          & 0.117          & 1.173             & 0.084          \\ 
    \midrule
    \multirow{4}{*}{Fair Baselines} & Projection-based CF ~\cite{rashid} & 0.419          & 0.244          & 0.674          & 0.307          & 0.407             & 0.030          \\
    & MF-$U_{abs}$ ~\cite{yao2017beyond}     & 0.163          & 0.007          & 0.627          & 0.184          & 0.629             & 0.026          \\
    & NCF via Resampling ~\cite{ekstrand2018all} & 0.315         & 0.153          & 0.586          & 0.222          & 0.442             & 0.025          \\
    & MF via Resampling ~\cite{ekstrand2018all} & 0.103          & 0.041          & 0.314          & 0.094          & 0.756             & 0.039          \\
    \bottomrule
\end{tabular}
}
	\caption{Comparison of proposed models with the baselines in career and college major recommendations on \emph{MovieLens} (17 \emph{careers}) and \emph{Facebook} (169 \emph{majors}). Higher is better for HR and NDCG; lower is better for $\epsilon_{mean}$ and $U_{abs}$. NFCF greatly improves fairness metrics and beats all baselines at recommendation except for NCF w Pre-train, a variant of NFCF without its fairness correction.
	\label{tab:quantitative_results}}
\end{table*}

%
\begin{table*}[t]
	\centering
	\small
	\resizebox{0.8\textwidth}{!}
	{
    \begin{tabular}{llll}
    \toprule
    \multicolumn{4}{c}{\emph{MovieLens} Dataset}\\
    \midrule
    \multicolumn{2}{c}{NFCF}                   & \multicolumn{2}{c}{NCF w/o Pre-train}      \\ \midrule
    Male                 & Female               & Male                 & Female               \\ \midrule
    college/grad student & college/grad student & sales/marketing      & customer service     \\
    executive/managerial & executive/managerial & academic/educator    & academic/educator    \\
    academic/educator    & technician/engineer  & executive/managerial & artist               \\
    technician/engineer  & academic/educator    & doctor/health care   & writer               \\
    programmer           & programmer           & college/grad student & college/grad student \\ \bottomrule
    &&&\\
    \toprule
    \multicolumn{4}{c}{\emph{Facebook} Dataset}\\
    \midrule
    \multicolumn{2}{c}{NFCF}                         & \multicolumn{2}{c}{NCF w/o Pre-train}         \\ \midrule
    Male                    & Female                  & Male                    & Female               \\ \midrule
    psychology              & psychology              & philosophy              & psychology           \\
    english literature      & english literature      & psychology              & nursing              \\
    graphic design          & music                   & computer science        & sociology            \\
    music                   & theatre                 & biochemistry            & graphic design       \\
    nursing                 & nursing                 & business admin.         & business marketing   \\
    liberal arts            & history                 & political science       & elementary education \\
    business admin.         & sociology               & business management     & cosmetology          \\
    biology                 & liberal arts            & medicine                & accounting           \\
    history                 & business admin.         & law                     & physical therapy     \\
    criminal justice        & biology                 & finance                 & music                \\ \bottomrule
    \end{tabular}
    }
	\caption{Top 5 (among 17 unique careers) and 10 (among 169 unique majors) most frequent career and college major recommendations on the \emph{MovieLens} and \emph{Facebook} datasets, respectively, to the overall male and female users using \emph{NFCF} and \emph{NCF w/o Pre-train} models . 
	\label{tab:qualitative}}
\end{table*}
\subsection{Experimental Settings}
All the models were trained via adaptive gradient descent optimization (Adam)  with learning rate = $0.001$ using pyTorch where we sampled $5$ negative instances per positive instance. The mini-batch size for all models was set to $2048$ and $256$ for user-page and user-career data, respectively, while the embedding size for users and items was set to $128$. The configuration of the DNN under \emph{NFCF} and \emph{NFCF\_embd} was $4$ hidden layers with $256$, $64$, $32$, $16$ neurons in each successive layer, ``relu'' and ``sigmoid'' activations for the hidden and output layers, respectively. We used the same DNN architecture for the \emph{NCF} and \emph{DNN Classifier} models. 

For the Facebook dataset, we held-out $1\%$ and $40\%$ from the user-page and user-college major data, respectively, as the test set, using the remainder for training.  
Since there are fewer users in the Movielens dataset, we held-out $1\%$ and $30\%$ from the user-movie and user-career data, respectively, as the test set, using the remainder for training. We further held-out $1\%$ and $20\%$ from the training user-nonsensitive item and user-sensitive item data, respectively, as the development set for each dataset. The tuning parameter $\lambda$ needs to be chosen as a trade-off between accuracy and fairness~\cite{foulds2018intersectional}. We chose $\lambda$ as $0.1$ for \emph{NFCF} and \emph{MF-$U_{abs}$} via a grid search on the development set according to similar criteria to ~\cite{foulds2018intersectional}, i.e. optimizing fairness while allowing up to $2\%$ degradation in accuracy from the typical model.

To evaluate the performance of item recommendation on the test data, since it is too time-consuming to rank all items for every user during evaluation~\cite{he2017neural}, we followed a common strategy in the literature ~\cite{elkahky2015multi}.  For non-sensitive items, we randomly sampled $100$ items which are not interacted by the user for each test instance, and ranked the test instance among the 100 items. For sensitive item recommendations, in the case of Facebook data we similarly randomly sampled $100$ college majors. For the MovieLens data, there are only $17$ unique careers, so we used the remaining $16$ careers when ranking the test instance.  
The performance of a ranked list is measured by the average Hit Ratio (HR) and Normalized Discounted Cumulative Gain (NDCG) ~\cite{he2015trirank}. The HR measures whether the test item is present in the top-$K$ list, while the NDCG accounts for the position of the hit by assigning higher scores to hits at top ranks. We calculated both metrics for each test user-item pair and reported the average score. 
Finally, we computed $\epsilon_{mean}$ and $U_{abs}$ on the test data for user-sensitive item to measure the fairness of the models in career and college major recommendations.

\subsection{Validation of NFCF Model Design} 
Before comparing to fair recommendation baseline models, we systematically validate our modeling choices for NFCF.

\textbf{Pre-training Task Performance:} 
We first study the performance for NCF and MF model at the pre-training task, \emph{Facebook page} and \emph{movie} recommendations (Table~\ref{tab:preTrain_performance}). NCF had substantially and consistently better performance compared to MF on the larger \emph{Facebook} dataset, and similar overall performance on \emph{MovieLens} (better in 2 of 4 metrics).

\textbf{Fine-Tuning Performance:} We fine-tuned these models on the interaction of users with the sensitive items for \emph{career} and \emph{college major} recommendations on \emph{MovieLens} and \emph{Facebook} dataset, respectively. Figure~\ref{fig:preTrain} shows top-$K$ recommendations from $17$ and $169$ unique \emph{careers} and \emph{college majors} using several ``typical'' baseline models that do not involve any fairness constraints, where $K$ ranges from $1$ to $5$ and $1$ to $25$ for MovieLens and Facebook dataset, respectively. \emph{NCF w/ Pre-train} had the best performance in HR and NDCG versus other baselines while our proposed \emph{NFCF} and \emph{NFCF\_embd} performed approximately similarly to \emph{NCF w/ Pre-train} for both datasets. Of the typical baselines, \emph{MF w/o Pre-train} and \emph{NCF w/o Pre-train} performed the second best for \emph{MovieLens} and \emph{Facebook} dataset, respectively. For the \emph{MovieLens} dataset, \emph{MF w/o Pre-train} performed better than \emph{MF w/ Pre-train}, presumably due to the relatively small dataset and having relatively few parameters to fine-tune, unlike for the DNN-based NCF model. 
\emph{BPMF} performed poorly despite using Bayesian inference for scarce data, perhaps due to \cite{salakhutdinov2008bayesian}'s initialization via older MF methods. 

\textbf{Visualization of Embedding De-biasing:} We visualized the PCA projections of the male and female vectors (Equation~\ref{eq:female}) before and after the linear projection-based de-biasing embeddings method, where PCA was performed based on all the embeddings.   
Figure~\ref{fig:debiasing_vectors} shows that the male and female vectors have very different directions and magnitudes. After de-biasing, the male and female vectors had a more similar direction and magnitude to each other.

\textbf{Ablation Study:} Finally, we conducted an ablation study in which the components of the method were removed one at a time. As shown in Table~\ref{tab:ablation_study}, 
there was a large degradation of the performance of \emph{NFCF} when pre-training was removed (de-biasing embeddings step was also removed, since there was no pre-trained user vector), or when NCF was replaced by MF.  Removing the de-biased embedding method lead to better HR and NDCG scores, but with an increase in the gender bias metrics. Similarly, learning without the fairness penalty lead to similar performance in HR and NDCG, but greatly increased gender bias. Therefore, \emph{\textbf{both} of the bias correction methods in the NFCF model are necessary to achieve the best level of fairness while maintaining a high recommendation accuracy}.

\subsection{Performance for Mitigating Gender Bias in Sensitive Item Recommendations}
We evaluated performance for career and college major recommendations in terms of accuracy (HR and NDCG) and fairness ($\epsilon_{mean}$ and $U_{abs}$). In Figure~\ref{fig:compare_fair_model}, we show that our proposed \emph{NFCF} and \emph{NFCF\_embd} models clearly outperformed all the fair baseline models in terms of HR and NDCG, regardless of the cut-off $K$. \emph{Projection-based CF} performed the second best on both datasets out of all the fair models.

In Table~\ref{tab:quantitative_results}, we show detailed results for the top $5$ and top $7$ recommendations on \emph{MovieLens} and for the top $10$ and top $25$ recommendations on the \emph{Facebook} dataset.  Our proposed \emph{NFCF} model was the most fair career and college major recommender in terms of $\epsilon_{mean}$, while our \emph{NFCF\_embd} was the most fair in terms of $U_{abs}$ on the \emph{Facebook} dataset. In the case of the \emph{MovieLens} dataset, our \emph{NFCF} model was the most fair recommender model in terms of both fairness metrics. \emph{NCF w/ Pre-train} performed best in the HR and NDCG metrics on both datasets. \emph{NFCF} and \emph{NFCF\_embd} 
had nearly as good HR and NDCG performance as \emph{NCF w/ Pre-train}, while also mitigating gender bias. We also found that the pre-training and fine-tuning approach reduced overfitting for \emph{NCF w/ Pre-train}, and thus improved the fairness metrics by reducing bias amplification.  This was not the case for \emph{MF w/ Pre-train}, presumably due to the limited number of pre-trained parameters to fine-tune.  
\emph{Projection-based CF}  and \emph{MF-$U_{abs}$} also showed relatively good performance in mitigating bias in terms of $U_{abs}$ compared to the typical models, but with a huge sacrifice in the accuracy. Similarly, \emph{NCF via Resampling}  and \emph{MF via Resampling} had poor performance in accuracy, but improved fairness to some extent over their corresponding ``typical'' models, \emph{NCF w/o Pre-train} and \emph{MF w/o Pre-train}, respectively.   

As a further qualitative experiment, we recommended the top-$1$ career and college major to each test male and female user via the \emph{NFCF} and \emph{NCF w/o Pre-train} models. In Table~\ref{tab:qualitative}, we show top $5$ and $10$ most frequent recommendations to the overall male and female users 
among the $17$ and $169$ unique careers and majors for \emph{MovieLens} and \emph{Facebook} dataset, respectively. \emph{NFCF} was found to recommend similar careers to both male and female users on average for both datasets, while \emph{NCF w/o Pre-train} encoded societal stereotypes in its recommendations. For example, \emph{NCF w/o Pre-train} recommends \emph{computer science} to male users and \emph{nursing} to female users on the \emph{Facebook} dataset while it recommends \emph{executive/managerial} to male users and \emph{customer service} to female users on the \emph{MovieLens} dataset.

\section{Conclusion}
We investigated gender bias in social-media based collaborative filtering. To address this problem, we introduced Neural Fair Collaborative Filtering (NFCF), a pre-training and fine-tuning method which corrects gender bias for recommending sensitive items such as careers or college majors with little loss in performance. On the \emph{MovieLens} and \emph{Facebook} datasets, we 
achieved better performance and fairness compared to an array of state-of-the-art models.

\bibliographystyle{coling}
\bibliography{references}

\end{document}